\documentclass[aps,prb,showpacs,twocolumn,longbibliography]{revtex4-1}
\usepackage[utf8]{inputenc}
\usepackage[colorlinks,pdfusetitle,urlcolor=blue,citecolor=blue,linkcolor=red,bookmarksnumbered,plainpages=false]{hyperref}
\usepackage{amsfonts}
\usepackage{amssymb}
\usepackage{amsmath}
\usepackage{graphicx}
\usepackage{epsfig}
\usepackage{subfigure}
\usepackage{appendix}

\begin{document}

\title{Ising chain with topological degeneracy induced by dissipation}
\author{K. L. Zhang}
\author{Z. Song}
\email{songtc@nankai.edu.cn}
\affiliation{School of Physics, Nankai University, Tianjin 300071, China}
\date{\today}

\begin{abstract}
The ground-state degeneracy of the quantum spin system is a characteristic of
nontrivial topology, when it is gapped and robust against disordered
perturbation. The corresponding quantum phase transition (QPT) is usually
driven by a real parameter. We study a non-Hermitian Ising chain with two
transverse fields, one real and another imaginary, based on the exact solution
and numerical simulation. We show that topological degeneracy still exists
and can be obtained by an imaginary transverse field from a topologically
trivial phase of a Hermitian system. The topological degeneracy is robust
against the random imaginary field and therefore expected to be immune to
disordered dissipation from the spontaneous decay in experiment. The
underlying mechanism is the nonlocal symmetry, which emerges only in
thermodynamic limit and relates two categories of QPTs in the quantum spin
system, rooted from topological order and symmetry breaking, respectively.
\end{abstract}

\maketitle

\section{Introduction}

Driving a quantum phase transition (QPT) is of interest to both condensed
matter physics and quantum information science. A varying parameter across
the critical point induces a symmetry spontaneous breaking for traditional
QPT \cite{sachdev1999quantum} and nonlocal topological order for topological QPT 
\cite{wen1990topological,wen2004origin}. In the recent works \cite{zhang2015topological,zhang2017majorana}, it turns out
that the local order parameter and the topological order parameter can coexist
to characterize the QPTs. The underlying mechanism of this fact is the
duality of the Kitaev model, which has been introduced to describe
one-dimensional spinless fermions with superconducting $p$-wave pairing \cite%
{kitaev2001unpaired}. On the one hand, It is the fermionized version of the familiar
one-dimensional ($1$D) transverse-field Ising model \cite{pfeuty1970one}, which is
one of the simplest solvable models exhibiting quantum criticality and
demonstrating a QPT with spontaneous symmetry breaking \cite{sachdev1999quantum}.
On the other hand, as the gene of a Kitaev model, its Majorana lattice is
the Su-Schrieffer-Heeger (SSH) model \cite{su1979solitons}, which has served as a
paradigmatic example of the $1$D system supporting topological character 
\cite{zak1989berry}. It manifests the typical feature of topological order since the
number of zero energy and edge states are immune to local perturbations \cite%
{asboth2016short}. The topological superconducting has been demonstrated by unpaired
Majorana modes exponentially localized at the ends of open Kitaev chains,
which are robust against disordered perturbation. A system with topological
phase can be a promising platform for quantum computation and information
processing due to the intrinsic stability of the topological feature \cite%
{nayak2008non,stern2010non,alicea2012new}.

So far, most of the investigations on the QPT are driven by varying a real
parameter at absolute zero temperature. However, in practice, a genuine
quantum system, such as cold atoms, is intrinsically non-Hermitian because of
spontaneous decay \cite{dalibard1992wave,dum1992monte,molmer1993monte,wiseman1996quantum,plenio1998quantum,lee2014heralded}.
On the other hand, a non-Hermitian Hamiltonian is no longer a forbidden regime
in quantum mechanics since the discovery that a certain class of
non-Hermitian Hamiltonians could exhibit entirely real spectra \cite{mostafazadeh2002pseudo,bender2002complex,bender1999pt,bender1998real}. It also turns out that certain types of
non-Hermitian terms may maintain the topological feature of the original
Hermitian system \cite{zhang2019topological}. A natural question is whether a
topologically trivial phase of a Hermitian system can be shifted to a
non-trivial phase with stable ground states by adding a
non-Hermitian term.

In this paper, we investigate a one-dimensional quantum Ising model with a 
complex transverse field. The aim of this paper is to study the consequence
of an imaginary transverse field on the topology of the ground state.
Intuitively, the imaginary transverse field may break the Hermiticity of the
Hamiltonian. However, it is shown that the non-Hermitian Ising model can be
mapped to a Hermitian Ising model in the context of a biorthogonal inner
product. This allows us to employ the same way for the Hermitian system, i.e.,
the robust degeneracy of ground states, to identify the nature of quantum
phases. We show that the topological degeneracy still exists in the presence
of a complex transverse field. We obtain the non-Hermitian version of the mapping
operator to connect two degenerate ground states in the topologically
nontrivial region. In addition, numerical simulation for a finite-size system
indicates the existence of topological degeneracy since the degeneracy
cannot be lifted by disordered perturbation on the imaginary transverse
field. Importantly, the Hamiltonian has a full real spectrum in
two different quantum phases. This property promises the stable ground
states, and allows the theoretical and experimental investigations for the
dynamical control and testing of the QPT.

This paper is organized as follows. In Sec. \ref{Hamiltonian and Hermitian
counterpart}, we present the model and its Hermitian counterpart. In Sec. %
\ref{Ground-state degeneracy and edge modes}, we analyze the topological
degeneracy in the Hermitian version. In Sec. \ref{Nonlocal symmetry and
robustness of degeneracy}, we propose the concept of nonlocal symmetry and
numerically study the robustness against the disordered imaginary field.
In Sec. \ref{Discussion}, we summarize the results and explore their
implications.

\section{Hamiltonian and Hermitian counterpart}

\label{Hamiltonian and Hermitian counterpart} We start our investigation by
considering a non-Hermitian Ising chain with a complex transverse field%
\begin{equation}
H=-\sum_{j=1}^{N-1}\sigma _{j}^{x}\sigma
_{j+1}^{x}+\sum_{j=1}^{N}g_{_{j}}\left( \sigma _{j}^{z}+i\gamma
_{_{j}}\sigma _{j}^{y}\right) ,  \label{H_com}
\end{equation}%
where $\sigma _{j}^{\lambda }$ ($\lambda =x,$ $y,$ $z$) are the Pauli
operators on site $j$, and $g_{_{j}}$ and $\gamma _{_{j}}$\ ($i=\sqrt{-1}$)
are real numbers. The non-Hermitian Hamiltonian $H$ has
pseudo-Hermiticity: either it has a real spectrum or else its complex
eigenvalues always come in complex conjugate pairs \cite{mostafazadeh2002pseudo}. As far as
we know, although non-Hermitian systems indeed have some peculiar features
and some of them have been proved to be equivalent to Hermitian
systems under some particular conditions \cite{dorey2001spectral, mostafazadeh2002pseudo3,mostafazadeh2004physical,mostafazadeh2003exact,jones2005pseudo,mostafazadeh2002pseudo2}%
, the complex field is always seen as unphysical. Recently some works,
including theoretical and experimental research on Lee Yang zeros, which are
the points on the complex plane of physical parameters, have been proposed \cite%
{wei2014phase,peng2015experimental,wei2015thermodynamic,ananikian2015imaginary,garcia2015density,chen2015near,krasnytska2015violation}. They relate a complex field to the
real world to some extent. On the other hand, it was proposed that \cite%
{lee2014heralded} an imaginary transverse field can be implemented by optically
pumping a qubit state into the auxiliary state with a scheme similar to
heralded entanglement protocols.

In order to explore the property of the non-Hermitian model, we introduce a
transformation 
\begin{eqnarray}
\tau _{j}^{x} &=&\sigma _{j}^{x},  \notag \\
\tau _{j}^{y} &=&\eta _{j}^{+}\sigma _{j}^{y}-i\eta _{j}^{-}\sigma _{j}^{z},
\notag \\
\tau _{j}^{z} &=&\eta _{j}^{+}\sigma _{j}^{z}+i\eta _{j}^{-}\sigma _{j}^{y},
\end{eqnarray}%
where the factors are $\eta _{j}^{+}=1/\sqrt{1-\gamma _{_{j}}^{2}}$ and $%
\eta _{j}^{-}=\gamma _{_{j}}/\sqrt{1-\gamma _{_{j}}^{2}}$. The new spin
operators still satisfy the Lie algebra commutation relations 
\begin{equation}
\left[ \tau _{j}^{\mu },\tau _{j}^{\nu }\right] =\sum_{\lambda
	=x,y,z}2i\epsilon ^{\mu \nu \lambda }\tau _{j}^{\lambda },
\label{NH_LieComm}
\end{equation}
although $\tau _{j}^{y}$\ and\ $\tau _{j}^{z}$\ are not Hermitian. Applying
the transformation on the Hamiltonian $H$, we have%
\begin{equation}
\mathcal{H}=-\sum_{j=1}^{N-1}\tau _{j}^{x}\tau
_{j+1}^{x}+\sum_{j=1}^{N}g_{_{j}}\sqrt{1-\gamma _{_{j}}^{2}}\tau _{j}^{z}.
\label{NH_Ising}
\end{equation}

Hamiltonian $\mathcal{H}$\ represents an Ising model with a real transverse
filed if $\left\vert \gamma _{_{j}}\right\vert <1$, and it has full real
spectrum although the spin operators $\tau _{j}^{y}$\ and $\tau _{j}^{z}$\
are not Hermitian. Within this region, $\mathcal{H}$\ shares the same
properties of $H$ with $\gamma _{_{j}}=0$,\ in the context of the biorthogonal
inner product. The connection between the eigenstates of $H$\ and $\mathcal{H%
}$\ can be obtained from the following relations%
\begin{equation}
\left\vert \pm \right\rangle _{_{j}}=\frac{1}{\sqrt{2}}\left( \xi ^{\pm
}\left\vert \uparrow \right\rangle _{_{j}}-\frac{\gamma }{\xi ^{\pm }}%
\left\vert \downarrow \right\rangle _{_{j}}\right) ,
\end{equation}%
where $\xi ^{\pm }=\sqrt{1\pm \sqrt{1-\gamma ^{2}}}$, and the corresponding
local spin vectors are defined as%
\begin{eqnarray}
\tau _{j}^{z}\left\vert \pm \right\rangle _{_{j}} &=&\pm \left\vert \pm
\right\rangle _{_{j}},  \notag \\
\sigma _{j}^{z}\left\vert \uparrow \right\rangle _{_{j}} &=&\left\vert
\uparrow \right\rangle _{_{j}},\sigma _{j}^{z}\left\vert \downarrow
\right\rangle _{_{j}}=-\left\vert \downarrow \right\rangle _{_{j}}.
\end{eqnarray}%
Accordingly, the biorthonomal complete set of eigenstates can be established
by the eigenstates of $\mathcal{H}^{\dag }$.

The aim of this paper is to investigate the effect of $\left\{ \gamma
_{_{j}}\right\} $\ on the topological feature of quantum phases.\ Now we
know that when an imaginary transverse field $\left\{ \gamma _{_{j}}\right\} 
$ is applied, the non-Hermitian Hamiltonian has a Hermitian counterpart,
which is the original Ising model with a real transverse field shifted by a $%
\gamma _{_{j}}$-related amount, i.e., $g_{i}\rightarrow g_{i}\sqrt{1-\gamma
_{_{j}}^{2}}$. In this sense, an imaginary field can lead to QPT. According to
quantum theory of QPT in a Hermitian system, a second-order QPT is
characterized by the divergence of the ground-state energy density. For the
Hamiltonian $H$ in Eq. (\ref{H_com}) with $g_{i}=g$ and $\gamma _{i}=\gamma
=0$, we have 
\begin{equation}
\lim_{g\rightarrow 1}\frac{\partial ^{2}\varepsilon _{\mathrm{g}}}{\partial
g^{2}}=\infty ,
\end{equation}%
where $\varepsilon _{\mathrm{g}}$\ is the density of the ground-state energy.
Meanwhile for the Hamiltonian $\mathcal{H}$ in Eq. (\ref{NH_Ising}) with $%
g_{i}=g$ and $\gamma _{i}=\gamma $, we have%
\begin{equation}
\frac{\partial \varepsilon _{\mathrm{g}}}{\partial \gamma }=-\frac{\gamma }{%
\sqrt{1-\gamma ^{2}}}\frac{\partial \varepsilon _{\mathrm{g}}}{\partial 
\sqrt{1-\gamma ^{2}}},
\end{equation}%
which results in%
\begin{equation}
\lim_{\gamma \rightarrow \gamma _{c}}\frac{\partial ^{2}\varepsilon _{%
\mathrm{g}}}{\partial \gamma ^{2}}=\infty ,
\end{equation}%
due to the replacement $g\rightarrow g\sqrt{1-\gamma ^{2}}$. It indicates
that $\gamma $\ drives a second-order QPT at $\gamma _{c}=\pm \sqrt{1-g^{-2}}
$. In the next two sections, we will analyze this issue from the topological
aspect.

\section{Ground-state degeneracy and edge modes}

\label{Ground-state degeneracy and edge modes}

We first revisit the connection between ground-state degeneracy and Majorana
edge modes in the Hermitian transverse field Ising chain in this section, and
then we extend the conclusion to the non-Hermitian version in the next section.
Consider the Hamiltonian $H$ in Eq. (\ref{H_com}) with $g_{i}=g$ and $\gamma
_{i}=\gamma =0$, 
\begin{equation}
H_{\text{spin}}=-\sum_{j=1}^{N-1}\sigma _{j}^{x}\sigma
_{j+1}^{x}+g\sum_{j=1}^{N}\sigma _{j}^{z},
\end{equation}%
which is a standard Ising model with an open boundary condition. The $2^{N}$-dimensional
complete set of basis can be constructed by applying operators $\sigma
_{j}^{+}=\left( \sigma _{j}^{x}+i\sigma _{j}^{y}\right) /2$ on a saturated
ferromagnetic state $\prod\nolimits_{j=1}^{N}\left\vert \downarrow
\right\rangle _{_{j}}$. The whole Hilbert space can be decomposed into two
invariant subspaces, with even and odd numbers of spin flips from the above
ferromagnetic state. Then all the eigenstates of $H_{\text{spin}}$\ can be
classified into two groups, denoted as $\{\left\vert \psi
_{j}^{+}\right\rangle \}$\ and $\{\left\vert \psi _{j}^{-}\right\rangle \}$,
respectively. In the following, we show the connection between two such 
groups of eigenstates, and the implication for topological degeneracy.

As a standard procedure, one can perform the Jordan-Wigner transformation \cite%
{jordan1993paulische}%
\begin{eqnarray}
\sigma _{j}^{x} &=&\prod\limits_{l<j}\left( 1-2c_{l}^{\dagger }c_{l}\right)
\left( c_{j}+c_{j}^{\dagger }\right),  \notag \\
\sigma _{j}^{y} &=&i\prod\limits_{l<j}\left( 1-2c_{l}^{\dagger }c_{l}\right)
\left( c_{j}-c_{j}^{\dagger }\right),  \notag \\
\sigma _{j}^{z} &=&2c_{j}^{\dag }c_{j}-1,
\end{eqnarray}%
to replace the Pauli operators by the fermion operators $c_{j}$. In this
paper, we focus on the system with an open boundary condition, in which there
are no differences between even and odd numbers of fermions. The Hamiltonian
is transformed to a Kitaev model%
\begin{eqnarray}
H_{\text{Kitaev}} &=&-\sum_{j=1}^{N-1}\left( c_{j}^{\dagger
}c_{j+1}+c_{j}^{\dag }c_{j+1}^{\dag }\right) +\text{\textrm{H.c.}}  \notag \\
&&+g\sum_{j=1}^{N}\left( 2c_{j}^{\dagger }c_{j}-1\right) .
\end{eqnarray}%
We note that the parity of particles number is conservative, i.e., $[\left(
-1\right) ^{\sum_{j=1}^{N}c_{j}^{\dagger }c_{j}},H_{\text{Kitaev}}]=0$,
corresponding to the classification of the eigenstates, $\left\{ \left\vert \psi
_{j}^{+}\right\rangle \right\} $\ and $\left\{ \left\vert \psi
_{j}^{-}\right\rangle \right\} $.

To get the solution of the model, one can introduce the Majorana
fermion operators $a_{j}=c_{j}^{\dagger }+c_{j}$ and $b_{j}=-i\left(
c_{j}^{\dagger }-c_{j}\right) $, and then obtain the Majorana representation
of the Hamiltonian, the core matrix of which is that of a $2N $-site SSH
chain in a single-particle invariant subspace. Based on the exact
diagonalization results of the SSH chain, the Hamiltonian $H_{\text{Kitaev}}$%
\ can be written as the diagonal form 
\begin{equation}
H_{\text{Kitaev}}=\sum_{n=1}^{N}\varepsilon _{n}(d_{n}^{\dagger }d_{n}-\frac{%
1}{2}).
\end{equation}%
Here $d_{n}$ is the fermonic operator, satisfying $\left\{ d_{n},d_{n^{\prime
}}\right\} =0,$ and $\left\{ d_{n},d_{n^{\prime }}^{\dag }\right\} =\delta
_{n,n^{\prime }}$. The spectrum $\varepsilon _{n}$\ and the explicit
expression of $d_{n}$ can be obtained by the diagonalization of the core
matrix:
\begin{eqnarray}
M_{\text{SSH}} &=&-\frac{1}{2}\sum_{j=1}^{N-1}\left\vert 2j\right\rangle
\left\langle 2j+1\right\vert  \notag \\
&&+\frac{1}{2}g\sum_{j=1}^{N}\left\vert 2j-1\right\rangle \left\langle
2j\right\vert +\text{\textrm{H.c.}}.  \label{M_SSH}
\end{eqnarray}%
On the other hand, no matter what the explicit solution is, we always have the
relations%
\begin{equation}
\left[ d_{n},H_{\text{Kitaev}}\right] =\varepsilon _{n}d_{n},\left[
d_{n}^{\dagger },H_{\text{Kitaev}}\right] =-\varepsilon _{n}d_{n}^{\dagger },
\end{equation}%
which result in the mapping between the eigenstates of $H_{\text{Kitaev}}$.
For an arbitrary eigenstate $\left\vert \psi \right\rangle $ of $H_{\text{%
Kitaev}}$ with the eigenenergy $E$, i.e.,%
\begin{equation}
H_{\text{Kitaev}}\left\vert \psi \right\rangle =E\left\vert \psi
\right\rangle ,
\end{equation}%
state $d_{n}\left\vert \psi \right\rangle $\ $\left( d_{n}^{\dag }\left\vert
\psi \right\rangle \right) $\ is also an eigenstate of $H_{\text{Kitaev}}$\
with the eigenenergy $E-\varepsilon _{n}$ $\left( E+\varepsilon _{n}\right) $%
, i.e.,%
\begin{equation}
H_{\text{Kitaev}}\left( d_{n}\left\vert \psi \right\rangle \right) =\left(
E-\varepsilon _{n}\right) \left( d_{n}\left\vert \psi \right\rangle \right) ,
\end{equation}%
and%
\begin{equation}
H_{\text{Kitaev}}\left( d_{n}^{\dag }\left\vert \psi \right\rangle \right)
=\left( E+\varepsilon _{n}\right) \left( d_{n}^{\dag }\left\vert \psi
\right\rangle \right) ,
\end{equation}%
if $d_{n}\left\vert \psi \right\rangle \neq 0$ $\left( d_{n}^{\dag
}\left\vert \psi \right\rangle \neq 0\right) $.

In this paper, we are interested in the topological degeneracy in the
topologically non-trivial phase, which arises from two zero-eigenenergy edge
states of the matrix in Eq. (\ref{M_SSH}) for $\left\vert g\right\vert <1$\
in the large $N$ limit. Actually, in small $N$, the energies of the
edge states are not exactly zero, but are exponentially small in $N$ \cite%
{asboth2016short}. However, here we are interested in the second-order QPT in the large $%
N $ limit, thus the terms that are exponentially small in $N $ are ignored
in the following discussion. It turns out that within the topological
region, the edge modes appear with $\varepsilon _{N}=0 $\ and the edge
operator $d_{N}$\ can be expressed as%
\begin{eqnarray}
d_{N} &=&\frac{1}{2}\sqrt{1-g^{2}}\sum_{j=1}^{N}[\left(
g^{j-1}+g^{N-j}\right) c_{j}^{\dagger }  \notag \\
&&+\left( g^{j-1}-g^{N-j}\right) c_{j}],
\end{eqnarray}%
i.e., $d_{N}$\ is a linear combination of particle and hole operators of
spinless fermions $c_{j}$\ on the edge. Furthermore, applying the inverse
Jordan-Wigner transformation, we can express $d_{N}$ as the combination of
spin operators:
\begin{eqnarray}
D_{N} &=&\frac{1}{2}\sqrt{1-g^{2}}\sum_{j=1}^{N}\prod\limits_{l<j}\left(
-\sigma _{l}^{z}\right) [\left( g^{j-1}+g^{N-j}\right) \sigma _{j}^{+} 
\notag \\
&&+\left( g^{j-1}-g^{N-j}\right) \sigma _{j}^{-}].
\end{eqnarray}%
In fact, $d_{N}$\ and $D_{N}$\ are identical, but only in different
representations. Obviously, from $[d_{N},H_{\text{Kitaev}}]=0$, we have 
\begin{equation}
\lbrack D_{N},H_{\text{spin}}]=[D_{N}^{\dag },H_{\text{spin}}]=0,
\label{symmetry}
\end{equation}%
which leads to the degeneracy of the eigenstates. The $D_{N}$\
operator can also be obtained by using the iterative method \cite{fendley2016strong,kemp2017long}. The properties of the fermion operator $D_{N}$ are similar to that of
the "strong edge zero mode" $\Psi $, which however, is a Majorana fermion
operator \cite{fendley2016strong,kemp2017long,else2017prethermal}. Moreover, the operator $D_{N}$ can be
obtained by the left and right "strong edge zero modes" $\Psi _{\mathrm{L/R}%
} $, i.e., $D_{N}=(\Psi _{\mathrm{L}}+i\Psi _{\mathrm{R}})/2$. We would
like to point out that the commutation relation in Eq. (\ref{symmetry}) can
be regarded as the symmetry of the system. Importantly, such a symmetry is
conditional, requiring $\left\vert g\right\vert <1$ in the large $N$ limit.
This accords with the symmetry breaking mechanism for QPT \cite{sachdev1999quantum}%
. On the other hand, the QPT also has topological characteristics since the
edge mode is robust against disorder perturbation. Here we do not review
this content in the Hermitian regime, but we investigate it directly in the
non-Hermitian regime in the next section.

\section{Nonlocal symmetry and robustness of degeneracy}

\label{Nonlocal symmetry and robustness of degeneracy} 
\begin{figure}[tbh]
\centering
\includegraphics[width=0.48\textwidth]{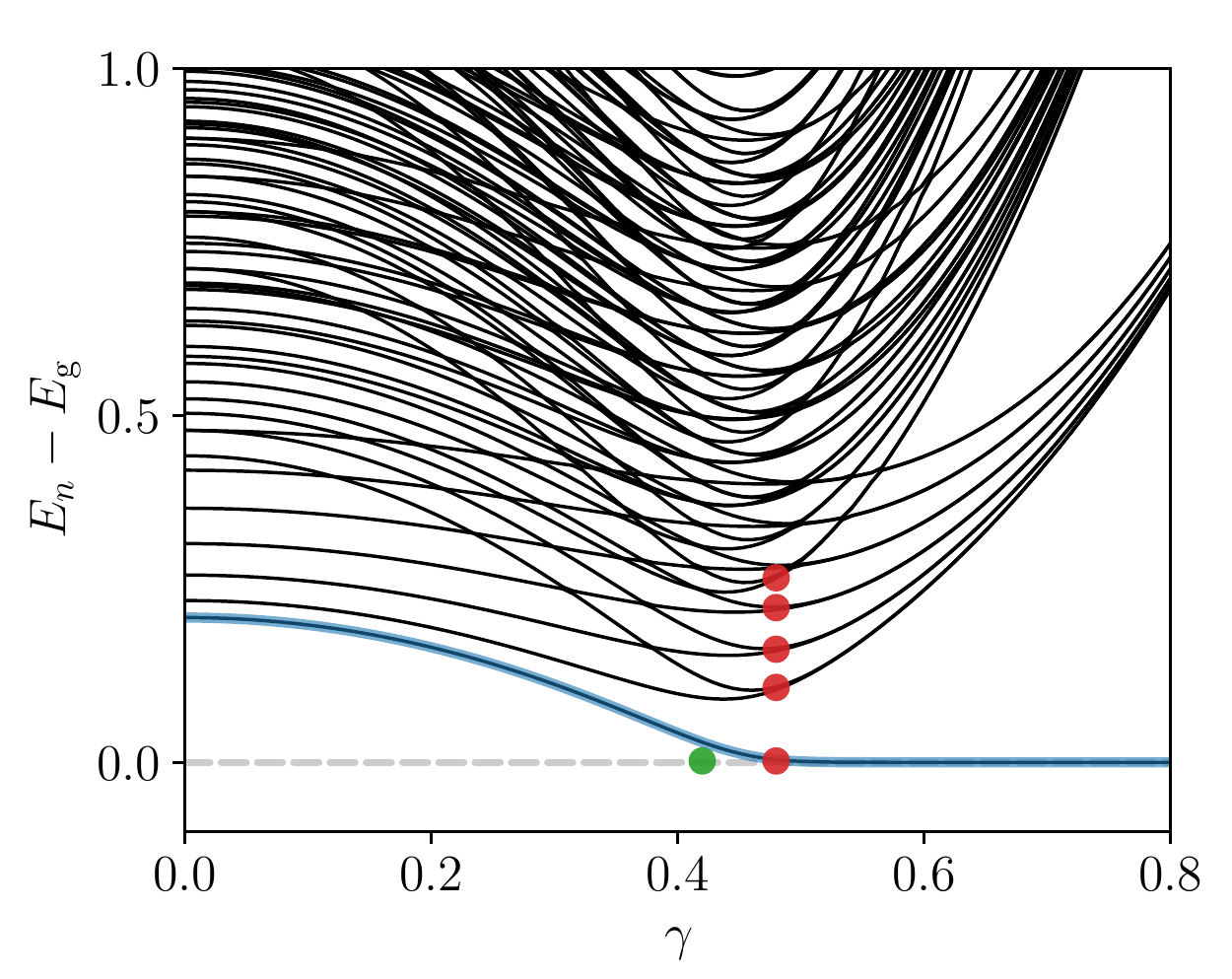}
\caption{The lower-energy levels (subtracted by the ground-state energy) of $%
\mathcal{H}$ as a function of $\protect\gamma $, obtained by exact
diagonalization numerically. Parameters of the system are $N=100$ and $g=1.1$%
. The energy gap between two ground states of two invariant subspaces, $%
\left\vert \protect\psi _{\mathrm{g}}^{+}\right\rangle $\ and $\left\vert 
\protect\psi _{\mathrm{g}}^{-}\right\rangle $,\ vanishes at the
pseudocritical point $\protect\gamma=0.48$, marked by a red dot, which is
closed to the critical point in the thermodynamic limit, $\protect\gamma _{c}=%
\protect\sqrt{1-g^{-2}}\approx0.42$, marked by a green dot. It indicates
that the imaginary transverse field can drive the second-order QPT.\ In
addition, we can see that there are many pairs of excited states, which also
turn to degenerate near the pseudo-critical point marked by the red dots. It
accords with the prediction obtained by the existence of nonlocal symmetry
in Eq. (\protect\ref{nonlocal symmetry}).}
\label{fig1}
\end{figure}

Starting from the
non-Hermitian Hamiltonian $H$ in Eq. (\ref{H_com}), it is tough to find out
the mapping operator along the same route in the last section \cite%
{Footnote,li2018topological}. However, we note that the commutation relation in Eq. (%
\ref{NH_LieComm}) is only based on the Lie algebra commutation relation of
spin operators $\left\{ \sigma _{j}^{\alpha }\right\} $ no matter whether they are
Hermitian or non-Hermitian. Then one can construct the mapping operator
directly by replacing $\left\{ \sigma _{j}^{\alpha }\right\} \ $with $%
\left\{ \tau _{j}^{\alpha }\right\} $. In parallel, we have the mapping
operator of the non-Hermitian version%
\begin{eqnarray}
\mathcal{D}_{N} &=&\frac{1}{2}\sqrt{1-\mathfrak{g}^{2}}\sum_{j=1}^{N}\prod%
\limits_{l<j}\left( -\tau _{l}^{z}\right) [\left( \mathfrak{g}^{j-1}+%
\mathfrak{g}^{N-j}\right) \tau _{j}^{+}  \notag \\
&&+\left( \mathfrak{g}^{j-1}-\mathfrak{g}^{N-j}\right) \tau _{j}^{-}],
\label{D_N}
\end{eqnarray}%
and its canonical conjugation 
\begin{eqnarray}
\overline{\mathcal{D}}_{N} &=&\frac{1}{2}\sqrt{1-\mathfrak{g}^{2}}%
\sum_{j=1}^{N}\prod\limits_{l<j}\left( -\tau _{l}^{z}\right) [\left( 
\mathfrak{g}^{j-1}+\mathfrak{g}^{N-j}\right) \tau _{j}^{-}  \notag \\
&&+\left( \mathfrak{g}^{j-1}-\mathfrak{g}^{N-j}\right) \tau _{j}^{+}],
\label{CD_N}
\end{eqnarray}%
in the large $N$ limit, where the equivalent real field $\mathfrak{g}=g\sqrt{%
1-\gamma ^{2}}$ and $\mathcal{D}_{N}\ (\overline{\mathcal{D}}_{N})$ is only
applicable within the region$\ \mathfrak{g}<1$, or $1>\gamma >\gamma _{c}$
with critical $\gamma _{c}=\sqrt{1-g^{-2}}$. When $\gamma \neq 0$%
, we note that $\mathcal{D}_{N}^{\dag }\neq \overline{\mathcal{D}}_{N}$ due
to the non-Hermitian nature of $\tau _{j}^{\pm }$. At $\gamma =1$, the
system is at the exceptional point, which is beyond our investigation.
Obviously, we still have%
\begin{equation}
\left[ \mathcal{D}_{N},\mathcal{H}\right] =\left[ \overline{\mathcal{D}}_{N},%
\mathcal{H}\right] =0  \label{nonlocal symmetry}
\end{equation}%
and the canonical commutation relations%
\begin{equation}
\left\{ \mathcal{D}_{N},\overline{\mathcal{D}}_{N}\right\} =1,\left( 
\mathcal{D}_{N}\right) ^{2}=\left( \overline{\mathcal{D}}_{N}\right) ^{2}=0,
\end{equation}%
which guarantee the existence of degeneracy of the eigenstates. Applying the
operators on the lowest energy eigenstates $\left\vert \psi _{\mathrm{g}%
}^{+}\right\rangle $\ and $\left\vert \psi _{\mathrm{g}}^{-}\right\rangle $
in two invariant subspaces, we have%
\begin{eqnarray}
\mathcal{D}_{N}\left\vert \psi _{\mathrm{g}}^{+}\right\rangle &=&\left\vert
\psi _{\mathrm{g}}^{-}\right\rangle ,\overline{\mathcal{D}}_{N}\left\vert
\psi _{\mathrm{g}}^{-}\right\rangle =\left\vert \psi _{\mathrm{g}%
}^{+}\right\rangle ,  \notag \\
\overline{\mathcal{D}}_{N}\left\vert \psi _{\mathrm{g}}^{+}\right\rangle &=&%
\mathcal{D}_{N}\left\vert \psi _{\mathrm{g}}^{-}\right\rangle =0,
\label{mapping}
\end{eqnarray}%
and then $\left\vert \psi _{\mathrm{g}}^{+}\right\rangle $\ and $\left\vert
\psi _{\mathrm{g}}^{-}\right\rangle $ are degenerate ground states. This
indicates that the existence of such a degeneracy depends on the value of $%
\gamma $\ through $\mathfrak{g}$. In Fig. \ref{fig1} we demonstrate this
feature by the plots of the lower-energy levels of $\mathcal{H}$\ on finite
size, as a function of $\gamma $. We can see that as $\gamma $ increases, the
energy gap closes at the pseudo-critical point, and states $\left\vert \psi
_{\mathrm{g}}^{+}\right\rangle $ and $\left\vert \psi _{\mathrm{g}%
}^{-}\right\rangle $ turn to degenerate. Meanwhile, many pairs of excited
states also become degenerate near the pseudocritical point, as expected in
Eq. (\ref{nonlocal symmetry}).

\begin{figure*}[tbh]
\centering
\includegraphics[width=1\textwidth]{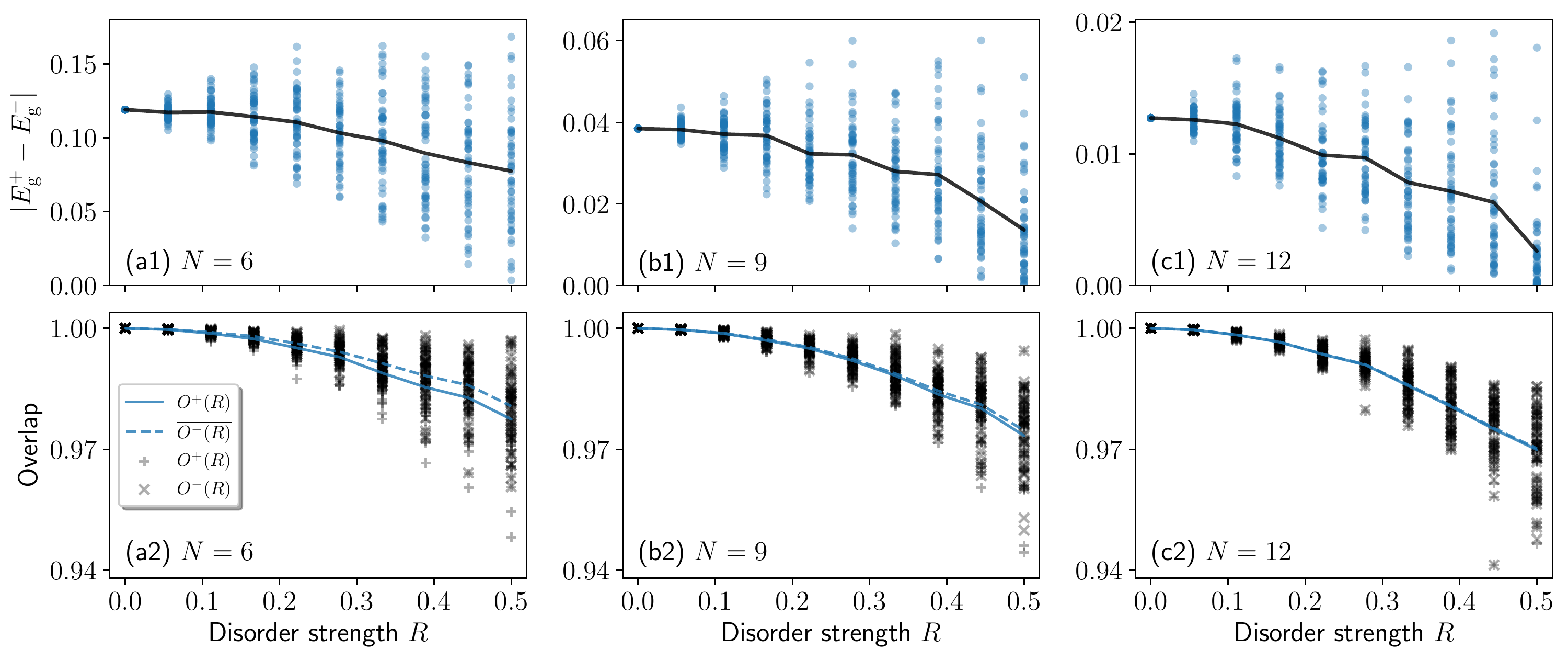}
\caption{Numerical results for the Hamiltonian in Eq. (\protect\ref{H_ran})
on finite $N$ chains obtained by exact diagonalization. Panels (a1)-(c1) and
(a2)-(c2) are energy gaps and the overlap $O^{\pm }(R)$ defined in Eq. (\protect
\ref{overlap}), respectively. For each disorder strength, numerical
simulations are performed $50$ times, whose average values are
represented by the lines. Other parameters are $g=0.8$ and $\protect\gamma %
=0.5$. The random strength $R$\ is taken under the condition of a full real
spectrum. Plots in panels (a1)-(c1) show that the energy gap of states $\left\vert 
\protect\psi _{\mathrm{g}}^{+}(R)\right\rangle$ and $\left\vert \protect\psi %
_{\mathrm{g}}^{-}(R)\right\rangle $ decreases as the size $N$ increases,
which accords with the prediction for large $N$. The average
values show that the energy gaps tend to become smaller as the random
strength increases up to $R=0.5$. Plots in panels (a2)-(c2) show that the overlap $%
O^{\pm }(R)$ decreases as the random strength $R$ increases. We find that
the degeneracy is robust against large disorder, and the ground state has no
evident deviation from that with zero $R$, corresponding to the uniform chain.}
\label{fig2}
\end{figure*}

Now we turn to investigate the performance of the ground states $\left\vert
\psi _{\mathrm{g}}^{\pm }\right\rangle $\ as the imaginary field is
disordered. Consider a model with random parameters%
\begin{equation}
\mathcal{H}_{\text{Ran}}=-\sum_{j=1}^{N-1}\sigma _{j}^{x}\sigma
_{j+1}^{x}+g\sum_{j=1}^{N}\left( \sigma _{j}^{z}+i\gamma _{_{j}}\sigma
_{j}^{y}\right) ,  \label{H_ran}
\end{equation}%
with the imaginary field being the form $\gamma _{j}=\gamma +\delta _{j}$,
where $\delta _{j}$\ is uniform random real numbers within the interval $%
\left( -R,R\right) $, taking the role of the disorder strength. We have the
mapping operator of the non-Hermitian version%
\begin{eqnarray}
\mathcal{D}_{N}(R) &=&\frac{1}{2}\sum_{j=1}^{N}\prod\limits_{l<j}\left(
-\tau _{l}^{z}\right) [\left( h_{j}^{+}+h_{j}^{-}\right) \tau _{j}^{+} 
\notag \\
&&+\left( h_{j}^{+}-h_{j}^{-}\right) \tau _{j}^{-}],
\end{eqnarray}%
and its canonical conjugation%
\begin{eqnarray}
\overline{\mathcal{D}}_{N}(R) &=&\frac{1}{2}\sum_{j=1}^{N}\prod\limits_{l<j}%
\left( -\tau _{l}^{z}\right) [\left( h_{j}^{+}+h_{j}^{-}\right) \tau _{j}^{-}
\notag \\
&&+\left( h_{j}^{+}-h_{j}^{-}\right) \tau _{j}^{+}],
\end{eqnarray}%
where%
\begin{eqnarray}
h_{j}^{+} &=&h_{1}^{+}\prod\limits_{m=1}^{j-1}\left( g\sqrt{1-\gamma _{m}^{2}%
}\right) ,  \notag \\
h_{j}^{-} &=&h_{N}^{-}\prod\limits_{m=j+1}^{N}\left( g\sqrt{1-\gamma _{m}^{2}%
}\right) ,
\end{eqnarray}
and $h_{1}^{+}$ ($h_{N}^{-}$) is determined by the normalization condition $%
\sum_{j=1}^{N}\left\vert h_{j}^{\pm }\right\vert ^{2}=1.$ These analytical
results can be obtained by the zero modes of the random core matrix
\begin{eqnarray}
M_{\text{Ran}} &=&-\frac{1}{2}\sum_{j=1}^{N-1}\left\vert 2j\right\rangle
\left\langle 2j+1\right\vert  \notag \\
&&+\frac{1}{2}g\sum_{j=1}^{N}\sqrt{1-\gamma _{j}^{2}}\left\vert
2j-1\right\rangle \left\langle 2j\right\vert +\text{\textrm{H.c.}}
\end{eqnarray}%
of the $2N$-site SSH chain in single-particle invariant subspace or by the iterative method \cite{fendley2016strong}. Here $\mathcal{D}_{N}(R)$ and 
$\overline{\mathcal{D}}_{N}(R)$\ are extensions of the operators in Eqs. (\ref%
{D_N}) and (\ref{CD_N}), i.e., $\mathcal{D}_{N}=\mathcal{D}_{N}(0)$\ and $%
\overline{\mathcal{D}}_{N}=\overline{\mathcal{D}}_{N}(0)$. The existence of
ground-state degeneracy [or the relations in Eq. (\ref{mapping})] in the
presence of nonzero $R$ depends on the existence of zero modes of the random
matrix $M_{\text{Ran}}$. More specifically, if the zero modes
exists, we can construct the mapping operators $\mathcal{D}_{N}(R)$ and $%
\overline{\mathcal{D}}_{N}(R)$ satisfying 
\begin{equation}
\left[ \mathcal{D}_{N}(R),\mathcal{H}_{\text{Ran}}\right] =\left[ \overline{%
\mathcal{D}}_{N}(R),\mathcal{H}_{\text{Ran}}\right] =0,
\label{nonlocal symmetry_R}
\end{equation}%
which is responsible for ground-state degeneracy. Here the existence of the
mapping operators that satisfy the above commutation relations is also know
as the normalizability of the operators \cite{fendley2016strong, kemp2017long}. It has been
shown that a slight disordered perturbation prevents the left of the zero
modes: in a SSH chain with non-uniform hoppings, if the average
intercell hopping is stronger than the average intracell hopping, the zero
modes exist in the large $N$ limit \cite{asboth2016short}.

Importantly,  the commutation relations in Eq. (\ref{nonlocal
symmetry_R}) can be regarded as the symmetries of the Hamiltonian $\mathcal{%
H}_{\text{Ran}}$. Three implications are worthy of discussion. (i) It is a
conditional symmetry, which requires the parameter $\left\vert g\right\vert 
\sqrt{1-\gamma _{j}^{2}}<1$. (ii) It only holds in the thermodynamic limit.
The two items accord with the mechanism of symmetry breaking for traditional
QPT. (iii) It is a nonlocal symmetry due to the string operator $%
\prod\nolimits_{l<j}\left( -\tau _{l}^{z}\right) $, and is robust against
the local disorder. It results in topological degeneracy as topological
order. In summary, the symmetries represented by commutation
relations Eq. (\ref{nonlocal symmetry_R}) relate two categories of QPTs in
the non-Hermitian or Hermitian quantum spin system, rooted from topological
order and symmetry breaking, respectively.

To demonstrate and verify the above analysis, finding out the extent of the
random strength to which the degeneracy can be lifted, we perform numerical
simulations for the original Hamiltonian in Eq. (\ref{H_ran}) on finite-size
chain. We focus on the deviations of the ground states arising from nonzero $%
R$. We compute the overlap between states $\left\vert \psi _{\mathrm{g}%
}^{\pm }\right\rangle $ with zero and nonzero $R$, which is defined as 
\begin{equation}
O^{\pm }(R)=\frac{\left\vert \langle \psi _{\mathrm{g}}^{\pm }(R)\left\vert
\psi _{\mathrm{g}}^{\pm }\right\rangle \right\vert }{\left\vert \left\vert
\psi _{\mathrm{g}}^{\pm }(R)\right\rangle \right\vert \left\vert \left\vert
\psi _{\mathrm{g}}^{\pm }\right\rangle \right\vert }.  \label{overlap}
\end{equation}%
In Fig. \ref{fig2}, we plot the energy gap and $O^{\pm }(R)$\ for a range of 
$R$, which ensures all the spectra to be real. The simulations are performed
by exact diagonalization of the matrices with random elements. The results
are influenced by the accident from the random number generator. The average
results from several random sequences show the following: (i) The energy gap
decreases as the size of the chain increases, which accords with the
prediction for large $N$. In the presence of the disordered field, the
quasidegeneracy is not lifted, but tends to become smaller as
the random strength increases up to $R=0.5$ (to ensure all $\gamma_{j}<1$),
within the real-spectrum region. (ii) The overlap $O^{\pm }(R)$ decreases as
the random strength $R$ increases. These indicate that degeneracy is robust
against large disorder, and the noise in the imaginary field does not affect
the ground states too much.

\section{Discussion}

\label{Discussion}

In this paper, we have studied the consequence of an imaginary transverse
field on the topological feature of a one-dimensional quantum Ising model.
The competition between the Ising interaction and the real transverse field
results in two different quantum phases with a full real spectrum, ordered and
disordered. We have shown that when an imaginary field is added, the
original disordered phase can be shifted to an ordered one. Although the
non-Hermitian Ising model cannot support directly a Majorana fermion
description, there still exists a symmetry-breaking mechanism under open
boundary conditions and the thermodynamic limit. It supports the topological
degeneracy due to its robustness in the presence of a random imaginary field.
Our work, including the numerical result for a small-size system, reveals that
disordered dissipation is constructive in establishing topological ground
states, which potentially can be utilized for developing inherently robust
artificial devices for topological quantum computation.
\\ \hspace*{\fill} \\

\acknowledgments This work was supported by the National Natural Science Foundation of China
(under Grant No. 11874225).


\begin{thebibliography}{43}%
	\makeatletter
	\providecommand \@ifxundefined [1]{%
		\@ifx{#1\undefined}
	}%
	\providecommand \@ifnum [1]{%
		\ifnum #1\expandafter \@firstoftwo
		\else \expandafter \@secondoftwo
		\fi
	}%
	\providecommand \@ifx [1]{%
		\ifx #1\expandafter \@firstoftwo
		\else \expandafter \@secondoftwo
		\fi
	}%
	\providecommand \natexlab [1]{#1}%
	\providecommand \enquote  [1]{``#1''}%
	\providecommand \bibnamefont  [1]{#1}%
	\providecommand \bibfnamefont [1]{#1}%
	\providecommand \citenamefont [1]{#1}%
	\providecommand \href@noop [0]{\@secondoftwo}%
	\providecommand \href [0]{\begingroup \@sanitize@url \@href}%
	\providecommand \@href[1]{\@@startlink{#1}\@@href}%
	\providecommand \@@href[1]{\endgroup#1\@@endlink}%
	\providecommand \@sanitize@url [0]{\catcode `\\12\catcode `\$12\catcode
		`\&12\catcode `\#12\catcode `\^12\catcode `\_12\catcode `\%12\relax}%
	\providecommand \@@startlink[1]{}%
	\providecommand \@@endlink[0]{}%
	\providecommand \url  [0]{\begingroup\@sanitize@url \@url }%
	\providecommand \@url [1]{\endgroup\@href {#1}{\urlprefix }}%
	\providecommand \urlprefix  [0]{URL }%
	\providecommand \Eprint [0]{\href }%
	\providecommand \doibase [0]{http://dx.doi.org/}%
	\providecommand \selectlanguage [0]{\@gobble}%
	\providecommand \bibinfo  [0]{\@secondoftwo}%
	\providecommand \bibfield  [0]{\@secondoftwo}%
	\providecommand \translation [1]{[#1]}%
	\providecommand \BibitemOpen [0]{}%
	\providecommand \bibitemStop [0]{}%
	\providecommand \bibitemNoStop [0]{.\EOS\space}%
	\providecommand \EOS [0]{\spacefactor3000\relax}%
	\providecommand \BibitemShut  [1]{\csname bibitem#1\endcsname}%
	\let\auto@bib@innerbib\@empty
	%</preamble>
	\bibitem [{\citenamefont {S.}(1999)}]{sachdev1999quantum}%
	\BibitemOpen
	\bibfield  {author} {\bibinfo {author} {\bibfnamefont {Subir}\ \bibnamefont
			{S.}},\ }\bibfield  {title} {\enquote {\bibinfo {title} {Quantum phase
				transitions},}\ }\href {\doibase 10.1088/2058-7058/12/4/23} {\bibfield
		{journal} {\bibinfo  {journal} {Physics World}\ }\textbf {\bibinfo {volume}
			{12}},\ \bibinfo {pages} {33--38} (\bibinfo {year} {1999})}\BibitemShut
	{NoStop}%
	\bibitem [{\citenamefont {Wen}(1990)}]{wen1990topological}%
	\BibitemOpen
	\bibfield  {author} {\bibinfo {author} {\bibfnamefont {X.~G.}\ \bibnamefont
			{Wen}},\ }\bibfield  {title} {\enquote {\bibinfo {title} {Topological orders
				in rigid states},}\ }\href {\doibase 10.1142/S0217979290000139} {\bibfield
		{journal} {\bibinfo  {journal} {Int. J. Mod. Phys. B}\ }\textbf {\bibinfo
			{volume} {04}},\ \bibinfo {pages} {239--271} (\bibinfo {year} {1990})},\
	\Eprint {http://arxiv.org/abs/https://doi.org/10.1142/S0217979290000139}
	{https://doi.org/10.1142/S0217979290000139} \BibitemShut {NoStop}%
	\bibitem [{\citenamefont {Wen}(2004)}]{wen2004origin}%
	\BibitemOpen
	\bibfield  {author} {\bibinfo {author} {\bibfnamefont {X.-G.}\ \bibnamefont
			{Wen}},\ }\href@noop {} {\emph {\bibinfo {title} {Quantum Field Theory of
				Many-Body Systems: From the Origin of Sound to an Origin of Light and
				Electrons}}}\ (\bibinfo  {publisher} {Oxford University Press, Oxford},\
	\bibinfo {year} {2004})\BibitemShut {NoStop}%
	\bibitem [{\citenamefont {Zhang}\ and\ \citenamefont
		{Song}(2015)}]{zhang2015topological}%
	\BibitemOpen
	\bibfield  {author} {\bibinfo {author} {\bibfnamefont {G.}~\bibnamefont
			{Zhang}}\ and\ \bibinfo {author} {\bibfnamefont {Z.}~\bibnamefont {Song}},\
	}\bibfield  {title} {\enquote {\bibinfo {title} {Topological characterization
				of extended quantum ising models},}\ }\href {\doibase
		10.1103/PhysRevLett.115.177204} {\bibfield  {journal} {\bibinfo  {journal}
			{Phys. Rev. Lett.}\ }\textbf {\bibinfo {volume} {115}},\ \bibinfo {pages}
		{177204} (\bibinfo {year} {2015})}\BibitemShut {NoStop}%
	\bibitem [{\citenamefont {Zhang}\ \emph {et~al.}(2017)\citenamefont {Zhang},
		\citenamefont {Li},\ and\ \citenamefont {Song}}]{zhang2017majorana}%
	\BibitemOpen
	\bibfield  {author} {\bibinfo {author} {\bibfnamefont {G.}~\bibnamefont
			{Zhang}}, \bibinfo {author} {\bibfnamefont {C.}~\bibnamefont {Li}}, \ and\
		\bibinfo {author} {\bibfnamefont {Z.}~\bibnamefont {Song}},\ }\bibfield
	{title} {\enquote {\bibinfo {title} {Majorana charges, winding numbers and
				chern numbers in quantum ising models},}\ }\href
	{https://doi.org/10.1038/s41598-017-08323-0} {\bibfield  {journal} {\bibinfo
			{journal} {Sci. Rep.}\ }\textbf {\bibinfo {volume} {7}},\ \bibinfo {pages}
		{1--13} (\bibinfo {year} {2017})}\BibitemShut {NoStop}%
	\bibitem [{\citenamefont {Kitaev}(2001)}]{kitaev2001unpaired}%
	\BibitemOpen
	\bibfield  {author} {\bibinfo {author} {\bibfnamefont {A~Y.}\ \bibnamefont
			{Kitaev}},\ }\bibfield  {title} {\enquote {\bibinfo {title} {Unpaired
				majorana fermions in quantum wires},}\ }\href {\doibase
		10.1070/1063-7869/44/10s/s29} {\bibfield  {journal} {\bibinfo  {journal}
			{Phys. Usp.}\ }\textbf {\bibinfo {volume} {44}},\ \bibinfo {pages} {131}
		(\bibinfo {year} {2001})}\BibitemShut {NoStop}%
	\bibitem [{\citenamefont {Pfeuty}(1970)}]{pfeuty1970one}%
	\BibitemOpen
	\bibfield  {author} {\bibinfo {author} {\bibfnamefont {P.}~\bibnamefont
			{Pfeuty}},\ }\bibfield  {title} {\enquote {\bibinfo {title} {The
				one-dimensional ising model with a transverse field},}\ }\href {\doibase
		https://doi.org/10.1016/0003-4916(70)90270-8} {\bibfield  {journal} {\bibinfo
			{journal} {Ann. Phys.}\ }\textbf {\bibinfo {volume} {57}},\ \bibinfo {pages}
		{79--90} (\bibinfo {year} {1970})}\BibitemShut {NoStop}%
	\bibitem [{\citenamefont {Su}\ \emph {et~al.}(1979)\citenamefont {Su},
		\citenamefont {Schrieffer},\ and\ \citenamefont {Heeger}}]{su1979solitons}%
	\BibitemOpen
	\bibfield  {author} {\bibinfo {author} {\bibfnamefont {W.~P.}\ \bibnamefont
			{Su}}, \bibinfo {author} {\bibfnamefont {J.~R.}\ \bibnamefont {Schrieffer}},
		\ and\ \bibinfo {author} {\bibfnamefont {A.~J.}\ \bibnamefont {Heeger}},\
	}\bibfield  {title} {\enquote {\bibinfo {title} {Solitons in
				polyacetylene},}\ }\href {\doibase 10.1103/PhysRevLett.42.1698} {\bibfield
		{journal} {\bibinfo  {journal} {Phys. Rev. Lett.}\ }\textbf {\bibinfo
			{volume} {42}},\ \bibinfo {pages} {1698} (\bibinfo {year}
		{1979})}\BibitemShut {NoStop}%
	\bibitem [{\citenamefont {Zak}(1989)}]{zak1989berry}%
	\BibitemOpen
	\bibfield  {author} {\bibinfo {author} {\bibfnamefont {J.}~\bibnamefont
			{Zak}},\ }\bibfield  {title} {\enquote {\bibinfo {title} {Berry’s phase for
				energy bands in solids},}\ }\href {\doibase 10.1103/PhysRevLett.62.2747}
	{\bibfield  {journal} {\bibinfo  {journal} {Phys. Rev. Lett.}\ }\textbf
		{\bibinfo {volume} {62}},\ \bibinfo {pages} {2747} (\bibinfo {year}
		{1989})}\BibitemShut {NoStop}%
	\bibitem [{\citenamefont {Asb{\'o}th}\ \emph {et~al.}(2016)\citenamefont
		{Asb{\'o}th}, \citenamefont {Oroszl{\'a}ny},\ and\ \citenamefont
		{P{\'a}lyi}}]{asboth2016short}%
	\BibitemOpen
	\bibfield  {author} {\bibinfo {author} {\bibfnamefont {J{\'a}nos~K}\
			\bibnamefont {Asb{\'o}th}}, \bibinfo {author} {\bibfnamefont
			{L{\'a}szl{\'o}}\ \bibnamefont {Oroszl{\'a}ny}}, \ and\ \bibinfo {author}
		{\bibfnamefont {Andr{\'a}s}\ \bibnamefont {P{\'a}lyi}},\ }\href@noop {}
	{\emph {\bibinfo {title} {A short course on topological insulators}}}\
	(\bibinfo  {publisher} {Springer},\ \bibinfo {year} {2016})\BibitemShut
	{NoStop}%
	\bibitem [{\citenamefont {Nayak}\ \emph {et~al.}(2008)\citenamefont {Nayak},
		\citenamefont {Simon}, \citenamefont {Stern}, \citenamefont {Freedman},\ and\
		\citenamefont {Sarma}}]{nayak2008non}%
	\BibitemOpen
	\bibfield  {author} {\bibinfo {author} {\bibfnamefont {C.}~\bibnamefont
			{Nayak}}, \bibinfo {author} {\bibfnamefont {S.~H}\ \bibnamefont {Simon}},
		\bibinfo {author} {\bibfnamefont {A.}~\bibnamefont {Stern}}, \bibinfo
		{author} {\bibfnamefont {M.}~\bibnamefont {Freedman}}, \ and\ \bibinfo
		{author} {\bibfnamefont {S.~D.}\ \bibnamefont {Sarma}},\ }\bibfield  {title}
	{\enquote {\bibinfo {title} {Non-abelian anyons and topological quantum
				computation},}\ }\href {https://doi.org/10.1103/RevModPhys.80.1083}
	{\bibfield  {journal} {\bibinfo  {journal} {Rev. Mod. Phys.}\ }\textbf
		{\bibinfo {volume} {80}},\ \bibinfo {pages} {1083} (\bibinfo {year}
		{2008})}\BibitemShut {NoStop}%
	\bibitem [{\citenamefont {Stern}(2010)}]{stern2010non}%
	\BibitemOpen
	\bibfield  {author} {\bibinfo {author} {\bibfnamefont {A.}~\bibnamefont
			{Stern}},\ }\bibfield  {title} {\enquote {\bibinfo {title} {Non-abelian
				states of matter},}\ }\href {https://doi.org/10.1038/nature08915} {\bibfield
		{journal} {\bibinfo  {journal} {Nature}\ }\textbf {\bibinfo {volume} {464}},\
		\bibinfo {pages} {187--193} (\bibinfo {year} {2010})}\BibitemShut {NoStop}%
	\bibitem [{\citenamefont {Alicea}(2012)}]{alicea2012new}%
	\BibitemOpen
	\bibfield  {author} {\bibinfo {author} {\bibfnamefont {J.}~\bibnamefont
			{Alicea}},\ }\bibfield  {title} {\enquote {\bibinfo {title} {New directions
				in the pursuit of majorana fermions in solid state systems},}\ }\href
	{\doibase 10.1088/0034-4885/75/7/076501} {\bibfield  {journal} {\bibinfo
			{journal} {Rep. Prog. Phys.}\ }\textbf {\bibinfo {volume} {75}},\ \bibinfo
		{pages} {076501} (\bibinfo {year} {2012})}\BibitemShut {NoStop}%
	\bibitem [{\citenamefont {Dalibard}\ \emph {et~al.}(1992)\citenamefont
		{Dalibard}, \citenamefont {Castin},\ and\ \citenamefont
		{M{\o}lmer}}]{dalibard1992wave}%
	\BibitemOpen
	\bibfield  {author} {\bibinfo {author} {\bibfnamefont {J.}~\bibnamefont
			{Dalibard}}, \bibinfo {author} {\bibfnamefont {Y.}~\bibnamefont {Castin}}, \
		and\ \bibinfo {author} {\bibfnamefont {K.}~\bibnamefont {M{\o}lmer}},\
	}\bibfield  {title} {\enquote {\bibinfo {title} {Wave-function approach to
				dissipative processes in quantum optics},}\ }\href {\doibase
		10.1103/PhysRevLett.68.580} {\bibfield  {journal} {\bibinfo  {journal} {Phys.
				Rev. Lett.}\ }\textbf {\bibinfo {volume} {68}},\ \bibinfo {pages} {580}
		(\bibinfo {year} {1992})}\BibitemShut {NoStop}%
	\bibitem [{\citenamefont {Dum}\ \emph {et~al.}(1992)\citenamefont {Dum},
		\citenamefont {Zoller},\ and\ \citenamefont {Ritsch}}]{dum1992monte}%
	\BibitemOpen
	\bibfield  {author} {\bibinfo {author} {\bibfnamefont {R.}~\bibnamefont
			{Dum}}, \bibinfo {author} {\bibfnamefont {P.}~\bibnamefont {Zoller}}, \ and\
		\bibinfo {author} {\bibfnamefont {H.}~\bibnamefont {Ritsch}},\ }\bibfield
	{title} {\enquote {\bibinfo {title} {Monte carlo simulation of the atomic
				master equation for spontaneous emission},}\ }\href {\doibase
		10.1103/PhysRevA.45.4879} {\bibfield  {journal} {\bibinfo  {journal} {Phys.
				Rev. A}\ }\textbf {\bibinfo {volume} {45}},\ \bibinfo {pages} {4879}
		(\bibinfo {year} {1992})}\BibitemShut {NoStop}%
	\bibitem [{\citenamefont {M{\o}lmer}\ \emph {et~al.}(1993)\citenamefont
		{M{\o}lmer}, \citenamefont {Castin},\ and\ \citenamefont
		{Dalibard}}]{molmer1993monte}%
	\BibitemOpen
	\bibfield  {author} {\bibinfo {author} {\bibfnamefont {K.}~\bibnamefont
			{M{\o}lmer}}, \bibinfo {author} {\bibfnamefont {Y.}~\bibnamefont {Castin}}, \
		and\ \bibinfo {author} {\bibfnamefont {J.}~\bibnamefont {Dalibard}},\
	}\bibfield  {title} {\enquote {\bibinfo {title} {Monte carlo wave-function
				method in quantum optics},}\ }\href {\doibase 10.1364/JOSAB.10.000524}
	{\bibfield  {journal} {\bibinfo  {journal} {J. Opt. Soc. Am. B}\ }\textbf
		{\bibinfo {volume} {10}},\ \bibinfo {pages} {524--538} (\bibinfo {year}
		{1993})}\BibitemShut {NoStop}%
	\bibitem [{\citenamefont {Wiseman}(1996)}]{wiseman1996quantum}%
	\BibitemOpen
	\bibfield  {author} {\bibinfo {author} {\bibfnamefont {H.~M.}\ \bibnamefont
			{Wiseman}},\ }\bibfield  {title} {\enquote {\bibinfo {title} {Quantum
				trajectories and quantum measurement theory},}\ }\href {\doibase
		10.1088/1355-5111/8/1/015} {\bibfield  {journal} {\bibinfo  {journal}
			{Quantum Semiclass. Opt.}\ }\textbf {\bibinfo {volume} {8}},\ \bibinfo
		{pages} {205} (\bibinfo {year} {1996})}\BibitemShut {NoStop}%
	\bibitem [{\citenamefont {Plenio}\ and\ \citenamefont
		{Knight}(1998)}]{plenio1998quantum}%
	\BibitemOpen
	\bibfield  {author} {\bibinfo {author} {\bibfnamefont {M.~B}\ \bibnamefont
			{Plenio}}\ and\ \bibinfo {author} {\bibfnamefont {P.~L}\ \bibnamefont
			{Knight}},\ }\bibfield  {title} {\enquote {\bibinfo {title} {The quantum-jump
				approach to dissipative dynamics in quantum optics},}\ }\href {\doibase
		10.1103/RevModPhys.70.101} {\bibfield  {journal} {\bibinfo  {journal} {Rev.
				Mod. Phys.}\ }\textbf {\bibinfo {volume} {70}},\ \bibinfo {pages} {101}
		(\bibinfo {year} {1998})}\BibitemShut {NoStop}%
	\bibitem [{\citenamefont {Lee}\ and\ \citenamefont
		{Chan}(2014)}]{lee2014heralded}%
	\BibitemOpen
	\bibfield  {author} {\bibinfo {author} {\bibfnamefont {T.~E}\ \bibnamefont
			{Lee}}\ and\ \bibinfo {author} {\bibfnamefont {C.-K.}\ \bibnamefont {Chan}},\
	}\bibfield  {title} {\enquote {\bibinfo {title} {Heralded magnetism in
				non-hermitian atomic systems},}\ }\href {\doibase 10.1103/PhysRevX.4.041001}
	{\bibfield  {journal} {\bibinfo  {journal} {Phys. Rev. X}\ }\textbf {\bibinfo
			{volume} {4}},\ \bibinfo {pages} {041001} (\bibinfo {year}
		{2014})}\BibitemShut {NoStop}%
	\bibitem [{\citenamefont
		{Mostafazadeh}(2002{\natexlab{a}})}]{mostafazadeh2002pseudo}%
	\BibitemOpen
	\bibfield  {author} {\bibinfo {author} {\bibfnamefont {A.}~\bibnamefont
			{Mostafazadeh}},\ }\bibfield  {title} {\enquote {\bibinfo {title}
			{Pseudo-hermiticity versus pt symmetry: the necessary condition for the
				reality of the spectrum of a non-hermitian hamiltonian},}\ }\href {\doibase
		10.1063/1.1418246} {\bibfield  {journal} {\bibinfo  {journal} {J. Math.
				Phys.}\ }\textbf {\bibinfo {volume} {43}},\ \bibinfo {pages} {205--214}
		(\bibinfo {year} {2002}{\natexlab{a}})}\BibitemShut {NoStop}%
	\bibitem [{\citenamefont {Bender}\ \emph {et~al.}(2002)\citenamefont {Bender},
		\citenamefont {Brody},\ and\ \citenamefont {Jones}}]{bender2002complex}%
	\BibitemOpen
	\bibfield  {author} {\bibinfo {author} {\bibfnamefont {C.~M}\ \bibnamefont
			{Bender}}, \bibinfo {author} {\bibfnamefont {D.~C}\ \bibnamefont {Brody}}, \
		and\ \bibinfo {author} {\bibfnamefont {H.~F}\ \bibnamefont {Jones}},\
	}\bibfield  {title} {\enquote {\bibinfo {title} {Complex extension of quantum
				mechanics},}\ }\href {\doibase 10.1103/PhysRevLett.89.270401} {\bibfield
		{journal} {\bibinfo  {journal} {Phys. Rev. Lett.}\ }\textbf {\bibinfo
			{volume} {89}},\ \bibinfo {pages} {270401} (\bibinfo {year}
		{2002})}\BibitemShut {NoStop}%
	\bibitem [{\citenamefont {Bender}\ \emph {et~al.}(1999)\citenamefont {Bender},
		\citenamefont {Boettcher},\ and\ \citenamefont {Meisinger}}]{bender1999pt}%
	\BibitemOpen
	\bibfield  {author} {\bibinfo {author} {\bibfnamefont {C.~M}\ \bibnamefont
			{Bender}}, \bibinfo {author} {\bibfnamefont {S.}~\bibnamefont {Boettcher}}, \
		and\ \bibinfo {author} {\bibfnamefont {P.~N}\ \bibnamefont {Meisinger}},\
	}\bibfield  {title} {\enquote {\bibinfo {title} {Pt-symmetric quantum
				mechanics},}\ }\href {https://doi.org/10.1063/1.532860} {\bibfield  {journal}
		{\bibinfo  {journal} {J. Math. Phys.}\ }\textbf {\bibinfo {volume} {40}},\
		\bibinfo {pages} {2201--2229} (\bibinfo {year} {1999})}\BibitemShut {NoStop}%
	\bibitem [{\citenamefont {Bender}\ and\ \citenamefont
		{Boettcher}(1998)}]{bender1998real}%
	\BibitemOpen
	\bibfield  {author} {\bibinfo {author} {\bibfnamefont {C.~M}\ \bibnamefont
			{Bender}}\ and\ \bibinfo {author} {\bibfnamefont {S.}~\bibnamefont
			{Boettcher}},\ }\bibfield  {title} {\enquote {\bibinfo {title} {Real spectra
				in non-hermitian hamiltonians having pt symmetry},}\ }\href {\doibase
		10.1103/PhysRevLett.80.5243} {\bibfield  {journal} {\bibinfo  {journal}
			{Phys. Rev. Lett.}\ }\textbf {\bibinfo {volume} {80}},\ \bibinfo {pages}
		{5243} (\bibinfo {year} {1998})}\BibitemShut {NoStop}%
	\bibitem [{\citenamefont {Zhang}\ \emph {et~al.}(2019)\citenamefont {Zhang},
		\citenamefont {Wu}, \citenamefont {Jin},\ and\ \citenamefont
		{Song}}]{zhang2019topological}%
	\BibitemOpen
	\bibfield  {author} {\bibinfo {author} {\bibfnamefont {K.~L.}\ \bibnamefont
			{Zhang}}, \bibinfo {author} {\bibfnamefont {H.~C.}\ \bibnamefont {Wu}},
		\bibinfo {author} {\bibfnamefont {L.}~\bibnamefont {Jin}}, \ and\ \bibinfo
		{author} {\bibfnamefont {Z.}~\bibnamefont {Song}},\ }\bibfield  {title}
	{\enquote {\bibinfo {title} {Topological phase transition independent of
				system non-hermiticity},}\ }\href {\doibase 10.1103/PhysRevB.100.045141}
	{\bibfield  {journal} {\bibinfo  {journal} {Phys. Rev. B}\ }\textbf {\bibinfo
			{volume} {100}},\ \bibinfo {pages} {045141} (\bibinfo {year}
		{2019})}\BibitemShut {NoStop}%
	\bibitem [{\citenamefont {Dorey}\ \emph {et~al.}(2001)\citenamefont {Dorey},
		\citenamefont {Dunning},\ and\ \citenamefont {Tateo}}]{dorey2001spectral}%
	\BibitemOpen
	\bibfield  {author} {\bibinfo {author} {\bibfnamefont {P.}~\bibnamefont
			{Dorey}}, \bibinfo {author} {\bibfnamefont {C.}~\bibnamefont {Dunning}}, \
		and\ \bibinfo {author} {\bibfnamefont {R.}~\bibnamefont {Tateo}},\ }\bibfield
	{title} {\enquote {\bibinfo {title} {Spectral equivalences, bethe ansatz
				equations, and reality properties in pt-symmetric quantum mechanics},}\
	}\href {\doibase 10.1088/0305-4470/34/28/305} {\bibfield  {journal} {\bibinfo
			{journal} {J Phys. A: Math Gen.}\ }\textbf {\bibinfo {volume} {34}},\
		\bibinfo {pages} {5679} (\bibinfo {year} {2001})}\BibitemShut {NoStop}%
	\bibitem [{\citenamefont
		{Mostafazadeh}(2002{\natexlab{b}})}]{mostafazadeh2002pseudo3}%
	\BibitemOpen
	\bibfield  {author} {\bibinfo {author} {\bibfnamefont {A.}~\bibnamefont
			{Mostafazadeh}},\ }\bibfield  {title} {\enquote {\bibinfo {title}
			{Pseudo-hermiticity versus pt-symmetry iii: Equivalence of pseudo-hermiticity
				and the presence of antilinear symmetries},}\ }\href
	{https://doi.org/10.1063/1.1489072} {\bibfield  {journal} {\bibinfo
			{journal} {J. Math. Phys.}\ }\textbf {\bibinfo {volume} {43}},\ \bibinfo
		{pages} {3944--3951} (\bibinfo {year} {2002}{\natexlab{b}})}\BibitemShut
	{NoStop}%
	\bibitem [{\citenamefont {Mostafazadeh}\ and\ \citenamefont
		{Batal}(2004)}]{mostafazadeh2004physical}%
	\BibitemOpen
	\bibfield  {author} {\bibinfo {author} {\bibfnamefont {A.}~\bibnamefont
			{Mostafazadeh}}\ and\ \bibinfo {author} {\bibfnamefont {A.}~\bibnamefont
			{Batal}},\ }\bibfield  {title} {\enquote {\bibinfo {title} {Physical aspects
				of pseudo-hermitian and pt-symmetric quantum mechanics},}\ }\href {\doibase
		10.1088/0305-4470/37/48/009} {\bibfield  {journal} {\bibinfo  {journal} {J.
				Phys. A: Math. Gen.}\ }\textbf {\bibinfo {volume} {37}},\ \bibinfo {pages}
		{11645} (\bibinfo {year} {2004})}\BibitemShut {NoStop}%
	\bibitem [{\citenamefont {Mostafazadeh}(2003)}]{mostafazadeh2003exact}%
	\BibitemOpen
	\bibfield  {author} {\bibinfo {author} {\bibfnamefont {A.}~\bibnamefont
			{Mostafazadeh}},\ }\bibfield  {title} {\enquote {\bibinfo {title} {Exact
				pt-symmetry is equivalent to hermiticity},}\ }\href {\doibase
		10.1088/0305-4470/36/25/312} {\bibfield  {journal} {\bibinfo  {journal} {J.
				Phys. A: Math Gen.}\ }\textbf {\bibinfo {volume} {36}},\ \bibinfo {pages}
		{7081} (\bibinfo {year} {2003})}\BibitemShut {NoStop}%
	\bibitem [{\citenamefont {Jones}(2005)}]{jones2005pseudo}%
	\BibitemOpen
	\bibfield  {author} {\bibinfo {author} {\bibfnamefont {H.~F.}\ \bibnamefont
			{Jones}},\ }\bibfield  {title} {\enquote {\bibinfo {title} {On
				pseudo-hermitian hamiltonians and their hermitian counterparts},}\ }\href
	{\doibase 10.1088/0305-4470/38/8/010} {\bibfield  {journal} {\bibinfo
			{journal} {J. Phys. A: Math Gen.}\ }\textbf {\bibinfo {volume} {38}},\
		\bibinfo {pages} {1741} (\bibinfo {year} {2005})}\BibitemShut {NoStop}%
	\bibitem [{\citenamefont
		{Mostafazadeh}(2002{\natexlab{c}})}]{mostafazadeh2002pseudo2}%
	\BibitemOpen
	\bibfield  {author} {\bibinfo {author} {\bibfnamefont {A.}~\bibnamefont
			{Mostafazadeh}},\ }\bibfield  {title} {\enquote {\bibinfo {title}
			{Pseudo-hermiticity versus pt-symmetry ii. a complete characterization of
				non-hermitian hamiltonians with a real spectrum},}\ }\href
	{https://doi.org/10.1063/1.1461427} {\bibfield  {journal} {\bibinfo
			{journal} {J. Math. Phys.}\ }\textbf {\bibinfo {volume} {43}},\ \bibinfo
		{pages} {2814--2816} (\bibinfo {year} {2002}{\natexlab{c}})}\BibitemShut
	{NoStop}%
	\bibitem [{\citenamefont {Wei}\ \emph {et~al.}(2014)\citenamefont {Wei},
		\citenamefont {Chen}, \citenamefont {Po},\ and\ \citenamefont
		{Liu}}]{wei2014phase}%
	\BibitemOpen
	\bibfield  {author} {\bibinfo {author} {\bibfnamefont {B.-B.}\ \bibnamefont
			{Wei}}, \bibinfo {author} {\bibfnamefont {S.-W.}\ \bibnamefont {Chen}},
		\bibinfo {author} {\bibfnamefont {H.-C.}\ \bibnamefont {Po}}, \ and\ \bibinfo
		{author} {\bibfnamefont {R.-B.}\ \bibnamefont {Liu}},\ }\bibfield  {title}
	{\enquote {\bibinfo {title} {Phase transitions in the complex plane of
				physical parameters},}\ }\href {https://doi.org/10.1038/srep05202} {\bibfield
		{journal} {\bibinfo  {journal} {Sci. Rep.}\ }\textbf {\bibinfo {volume}
			{4}},\ \bibinfo {pages} {5202} (\bibinfo {year} {2014})}\BibitemShut
	{NoStop}%
	\bibitem [{\citenamefont {Peng}\ \emph {et~al.}(2015)\citenamefont {Peng},
		\citenamefont {Zhou}, \citenamefont {Wei}, \citenamefont {Cui}, \citenamefont
		{Du},\ and\ \citenamefont {Liu}}]{peng2015experimental}%
	\BibitemOpen
	\bibfield  {author} {\bibinfo {author} {\bibfnamefont {X.}~\bibnamefont
			{Peng}}, \bibinfo {author} {\bibfnamefont {H.}~\bibnamefont {Zhou}}, \bibinfo
		{author} {\bibfnamefont {B.-B.}\ \bibnamefont {Wei}}, \bibinfo {author}
		{\bibfnamefont {J.}~\bibnamefont {Cui}}, \bibinfo {author} {\bibfnamefont
			{J.}~\bibnamefont {Du}}, \ and\ \bibinfo {author} {\bibfnamefont {R.-B.}\
			\bibnamefont {Liu}},\ }\bibfield  {title} {\enquote {\bibinfo {title}
			{Experimental observation of lee-yang zeros},}\ }\href {\doibase
		10.1103/PhysRevLett.114.010601} {\bibfield  {journal} {\bibinfo  {journal}
			{Phys. Rev. Lett.}\ }\textbf {\bibinfo {volume} {114}},\ \bibinfo {pages}
		{010601} (\bibinfo {year} {2015})}\BibitemShut {NoStop}%
	\bibitem [{\citenamefont {Wei}\ \emph {et~al.}(2015)\citenamefont {Wei},
		\citenamefont {Jiang},\ and\ \citenamefont {Liu}}]{wei2015thermodynamic}%
	\BibitemOpen
	\bibfield  {author} {\bibinfo {author} {\bibfnamefont {B.-B.}\ \bibnamefont
			{Wei}}, \bibinfo {author} {\bibfnamefont {Z.-F.}\ \bibnamefont {Jiang}}, \
		and\ \bibinfo {author} {\bibfnamefont {R.-B.}\ \bibnamefont {Liu}},\
	}\bibfield  {title} {\enquote {\bibinfo {title} {Thermodynamic holography},}\
	}\href {https://doi.org/10.1038/srep15077} {\bibfield  {journal} {\bibinfo
			{journal} {Sci. Rep.}\ }\textbf {\bibinfo {volume} {5}},\ \bibinfo {pages}
		{15077} (\bibinfo {year} {2015})}\BibitemShut {NoStop}%
	\bibitem [{\citenamefont {Ananikian}\ and\ \citenamefont
		{Kenna}(2015)}]{ananikian2015imaginary}%
	\BibitemOpen
	\bibfield  {author} {\bibinfo {author} {\bibfnamefont {N.}~\bibnamefont
			{Ananikian}}\ and\ \bibinfo {author} {\bibfnamefont {R.}~\bibnamefont
			{Kenna}},\ }\bibfield  {title} {\enquote {\bibinfo {title} {Imaginary
				magnetic fields in the real world},}\ }\href {\doibase 10.1103/Physics.8.2}
	{\bibfield  {journal} {\bibinfo  {journal} {Physics}\ }\textbf {\bibinfo
			{volume} {8}},\ \bibinfo {pages} {2} (\bibinfo {year} {2015})}\BibitemShut
	{NoStop}%
	\bibitem [{\citenamefont {Garc{\'\i}a-Saez}\ and\ \citenamefont
		{Wei}(2015)}]{garcia2015density}%
	\BibitemOpen
	\bibfield  {author} {\bibinfo {author} {\bibfnamefont {A.}~\bibnamefont
			{Garc{\'\i}a-Saez}}\ and\ \bibinfo {author} {\bibfnamefont {T.-C.}\
			\bibnamefont {Wei}},\ }\bibfield  {title} {\enquote {\bibinfo {title}
			{Density of yang-lee zeros in the thermodynamic limit from tensor network
				methods},}\ }\href {\doibase 10.1103/PhysRevB.92.125132} {\bibfield
		{journal} {\bibinfo  {journal} {Phys. Rev. B}\ }\textbf {\bibinfo {volume}
			{92}},\ \bibinfo {pages} {125132} (\bibinfo {year} {2015})}\BibitemShut
	{NoStop}%
	\bibitem [{\citenamefont {Chen}\ \emph {et~al.}(2015)\citenamefont {Chen},
		\citenamefont {Wu}, \citenamefont {Zhang},\ and\ \citenamefont
		{Rabitz}}]{chen2015near}%
	\BibitemOpen
	\bibfield  {author} {\bibinfo {author} {\bibfnamefont {Q.-M.}\ \bibnamefont
			{Chen}}, \bibinfo {author} {\bibfnamefont {R.-B.}\ \bibnamefont {Wu}},
		\bibinfo {author} {\bibfnamefont {T.-M.}\ \bibnamefont {Zhang}}, \ and\
		\bibinfo {author} {\bibfnamefont {H.}~\bibnamefont {Rabitz}},\ }\bibfield
	{title} {\enquote {\bibinfo {title} {Near-time-optimal control for quantum
				systems},}\ }\href {\doibase 10.1103/PhysRevA.92.063415} {\bibfield
		{journal} {\bibinfo  {journal} {Phys. Rev. A}\ }\textbf {\bibinfo {volume}
			{92}},\ \bibinfo {pages} {063415} (\bibinfo {year} {2015})}\BibitemShut
	{NoStop}%
	\bibitem [{\citenamefont {Krasnytska}\ \emph {et~al.}(2015)\citenamefont
		{Krasnytska}, \citenamefont {Berche}, \citenamefont {Holovatch},\ and\
		\citenamefont {Kenna}}]{krasnytska2015violation}%
	\BibitemOpen
	\bibfield  {author} {\bibinfo {author} {\bibfnamefont {M}~\bibnamefont
			{Krasnytska}}, \bibinfo {author} {\bibfnamefont {B.}~\bibnamefont {Berche}},
		\bibinfo {author} {\bibfnamefont {Y.}~\bibnamefont {Holovatch}}, \ and\
		\bibinfo {author} {\bibfnamefont {R.}~\bibnamefont {Kenna}},\ }\bibfield
	{title} {\enquote {\bibinfo {title} {Violation of lee-yang circle theorem for
				ising phase transitions on complex networks},}\ }\href {\doibase
		10.1209/0295-5075/111/60009} {\bibfield  {journal} {\bibinfo  {journal} {EPL
				(Europhysics Letters)}\ }\textbf {\bibinfo {volume} {111}},\ \bibinfo {pages}
		{60009} (\bibinfo {year} {2015})}\BibitemShut {NoStop}%
	\bibitem [{\citenamefont {Jordan}\ and\ \citenamefont
		{Wigner}(1993)}]{jordan1993paulische}%
	\BibitemOpen
	\bibfield  {author} {\bibinfo {author} {\bibfnamefont {P.}~\bibnamefont
			{Jordan}}\ and\ \bibinfo {author} {\bibfnamefont {E.~P.}\ \bibnamefont
			{Wigner}},\ }\enquote {\bibinfo {title} {{\"u}ber das paulische
			{\"a}quivalenzverbot},}\ in\ \href {\doibase 10.1007/978-3-662-02781-3_9}
	{\emph {\bibinfo {booktitle} {The Collected Works of Eugene Paul Wigner}}}\
	(\bibinfo  {publisher} {Springer},\ \bibinfo {year} {1993})\ pp.\ \bibinfo
	{pages} {109--129}\BibitemShut {NoStop}%
	\bibitem [{\citenamefont {Fendley}(2016)}]{fendley2016strong}%
	\BibitemOpen
	\bibfield  {author} {\bibinfo {author} {\bibfnamefont {P.}~\bibnamefont
			{Fendley}},\ }\bibfield  {title} {\enquote {\bibinfo {title} {Strong zero
				modes and eigenstate phase transitions in the xyz/interacting majorana
				chain},}\ }\href {\doibase 10.1088/1751-8113/49/30/30lt01} {\bibfield
		{journal} {\bibinfo  {journal} {J. Phys. A: Math. Theo.}\ }\textbf {\bibinfo
			{volume} {49}},\ \bibinfo {pages} {30LT01} (\bibinfo {year}
		{2016})}\BibitemShut {NoStop}%
	\bibitem [{\citenamefont {Kemp}\ \emph {et~al.}(2017)\citenamefont {Kemp},
		\citenamefont {Yao}, \citenamefont {Laumann},\ and\ \citenamefont
		{Fendley}}]{kemp2017long}%
	\BibitemOpen
	\bibfield  {author} {\bibinfo {author} {\bibfnamefont {J.}~\bibnamefont
			{Kemp}}, \bibinfo {author} {\bibfnamefont {N.~Y}\ \bibnamefont {Yao}},
		\bibinfo {author} {\bibfnamefont {C.~R}\ \bibnamefont {Laumann}}, \ and\
		\bibinfo {author} {\bibfnamefont {P.}~\bibnamefont {Fendley}},\ }\bibfield
	{title} {\enquote {\bibinfo {title} {Long coherence times for edge spins},}\
	}\href {\doibase 10.1088/1742-5468/aa73f0} {\bibfield  {journal} {\bibinfo
			{journal} {J. Stat. Mech.}\ }\textbf {\bibinfo {volume} {2017}},\ \bibinfo
		{pages} {063105} (\bibinfo {year} {2017})}\BibitemShut {NoStop}%
	\bibitem [{\citenamefont {Else}\ \emph {et~al.}(2017)\citenamefont {Else},
		\citenamefont {Fendley}, \citenamefont {Kemp},\ and\ \citenamefont
		{Nayak}}]{else2017prethermal}%
	\BibitemOpen
	\bibfield  {author} {\bibinfo {author} {\bibfnamefont {D.~V}\ \bibnamefont
			{Else}}, \bibinfo {author} {\bibfnamefont {P.}~\bibnamefont {Fendley}},
		\bibinfo {author} {\bibfnamefont {J.}~\bibnamefont {Kemp}}, \ and\ \bibinfo
		{author} {\bibfnamefont {C.}~\bibnamefont {Nayak}},\ }\bibfield  {title}
	{\enquote {\bibinfo {title} {Prethermal strong zero modes and topological
				qubits},}\ }\href {\doibase 10.1103/PhysRevX.7.041062} {\bibfield  {journal}
		{\bibinfo  {journal} {Phys. Rev. X}\ }\textbf {\bibinfo {volume} {7}},\
		\bibinfo {pages} {041062} (\bibinfo {year} {2017})}\BibitemShut {NoStop}%
	\bibitem{Footnote} Actually, one can develop a similar formation of Majorana
	fermion for the non-Hermitian spin operators $\left\{ \tau_{j}^{\alpha
	}\right\}$. Here the Majorana fermion is no longer self-conjugate particle 
	\cite{li2018topological}.
	\bibitem [{\citenamefont {Li}\ \emph {et~al.}(2018)\citenamefont {Li},
		\citenamefont {Zhang}, \citenamefont {Zhang},\ and\ \citenamefont
		{Song}}]{li2018topological}%
	\BibitemOpen
	\bibfield  {author} {\bibinfo {author} {\bibfnamefont {C.}~\bibnamefont
			{Li}}, \bibinfo {author} {\bibfnamefont {X.~Z.}\ \bibnamefont {Zhang}},
		\bibinfo {author} {\bibfnamefont {G.}~\bibnamefont {Zhang}}, \ and\ \bibinfo
		{author} {\bibfnamefont {Z.}~\bibnamefont {Song}},\ }\bibfield  {title}
	{\enquote {\bibinfo {title} {Topological phases in a kitaev chain with
				imbalanced pairing},}\ }\href {\doibase 10.1103/PhysRevB.97.115436}
	{\bibfield  {journal} {\bibinfo  {journal} {Phys. Rev. B}\ }\textbf {\bibinfo
			{volume} {97}},\ \bibinfo {pages} {115436} (\bibinfo {year}
		{2018})}\BibitemShut {NoStop}%
\end{thebibliography}
\end{document}